\documentclass[10pt,a4paper]{article}
\usepackage{amsmath}
\usepackage{amssymb}
\def\emline#1#2#3#4#5#6{%
       \put(#1,#2){\special{em:moveto}}%
       \put(#4,#5){\special{em:lineto}}}

\def\beq{\begin{equation}}
\def\eeq{\end{equation}}

\def\be{\begin{displaymath}}
\def\ee{\end{displaymath}}

\newtheorem{thm}{Theorem}

\def\p{\partial}
\def\pb{\bar{\partial}}
\def\dir{{\CD}\kern -4pt \slash \quad}
\def\Det{{\rm Det}}

\def\Tr{\mathop{\rm Tr}\nolimits}

\def\rank{\mathop{\rm rank}\nolimits}
\def\Ker{\mathop{\rm Ker}\nolimits}
\def\dim{\mathop{\rm dim}\nolimits}
\def\End{\mathop{\rm End}\nolimits}
\def\Hilb{\mathop{\rm Hilb}\nolimits}
\def\rep{\mathop{\rm rep}\nolimits}
\def\Hom{\mathop{\rm Hom}\nolimits}
\def\Image{\mathop{\rm Image}\nolimits}
\def\lhd{\triangleleft}
\def\vf{{\varphi}}
\newcommand{\R}{{\mathbb R}}
\newcommand{\C}{{\mathbb C}}
\newcommand{\A}{{\mathbb A}}
\newcommand{\M}{{\mathbb M}}

\newcommand{\CP}[1]{\C P^{#1}}
\renewcommand{\d}{\mathrm{d}}
\def\a{\alpha}
\def\b{\beta}
\def\l{\lambda}
\def\s{\sigma}
\def\z{\zeta}

\def\dd{{\rm d}}
\def\bx{{\bf G}}

\def\tb{\bar{t}}
\def\zb{\bar{z}}

\def\mapright#1{\smash{\mathop{\longrightarrow}\limits^{#1}}}

\def\asd{{anti-self-dual }}

\def\CD{{\cal D}}
\def\CE{{\cal E}}
\def\CF{{\cal F}}

\def\CI{{\cal I}}

\def\CS{{\cal S}}

\def\CU{{\cal U}}

\def\SWrefs{\cite{GKMMM,DW,MW,BMMM,BMMM1,BMMM2,BM,BK,Mar,GM}}

\title{
Instantons, Hilbert Schemes and Integrability\thanks{Based on
talks given by H.W.B.~at the Nato Advanced Research Workshops:
``Integrable Hierarchies and Modern Physical Theories'',  Chicago,
July 2000; ``Dynamical Symmetries of Integrable Quantum Field
Theories", Kiev, September 2000. } }

\author{H. W. Braden$^{1}$\thanks{hwb@ed.ac.uk}, N. A. Nekrasov$^{2,3}$\thanks{nikita@ihes.fr} \
\\ \normalsize \em $^{1}$Department of Mathematics and
Statistics,  The University of Edinburgh,  Edinburgh, UK
\\
\normalsize \em $^{2}$Institut des Hautes Etudes Scientifiques,
Bures-sur-Yvette, France\\ \normalsize \em $^{3}$Institute for
Theoretical and Experimental Physics, Moscow, Russia }

\date{}

\begin{document}

\renewcommand{\thepage}{}

\maketitle

\begin{abstract}
We review the deformed instanton equations making connection with
Hilbert schemes and integrable systems. A single $U(1)$  instanton
is shown to be \asd\ with respect to the Burns metric.

\end{abstract}



\section{Introduction}
The aim of the present review is to describe various settings surrounding
the matrix equations
\begin{align}\label{cmom}
[B_2,B_1]+IJ &= \ {\zeta}_{c} {\bf 1}_{V},
\\
[B_1,B_1^{\dagger}]+[B_2,B_2^{\dagger}]+I I\sp\dagger-J\sp\dagger J &=
2\zeta_{r} {\bf 1}_{V},\label{rmom}
\end{align}
where $B_{1,2}\in M_v(\C)$, $
I\in M_{v\times w}(\C)$, $ J\in M_{w\times v}(\C)$.
We will be interested in the space of solutions to these equations up to
equivalence under an action of $GL(\alpha):=GL(v,\C)\times GL(w,\C)$.
These equations arise naturally in the context of integrable systems
which will be recalled in the next section. The space of such matrices
describes the phase space of the integrable system  and we will refer
to it as a ``moduli space" as it describes the system at all energies and
momenta. This space carries a natural hyper-K\"ahler
structure and possesses several moment maps which are also reviewed.

Now the same moduli space also describes other interesting
phenomena. Indeed, if there is to be a simple motto describing
this talk it is: ``Phase spaces of completely integrable systems
give interesting moduli spaces for field theories". This is well
known in the context of Seiberg-Witten theory \cite{BK,Mar} but
true more generally \cite{RS,BH,BS}. In the present setting the
same moduli space parametrises the (semistable) torsion free
sheaves on $\CP2$ whose restriction on the projective line
${\ell}_{\infty}$ at infinity is trivial, as was shown first by
Nakajima. The connection between the Calogero-Moser systems and
instanton/sheaf moduli was noted by Nekrasov \cite{disser} and
Wilson \cite{wilson}. When the right-hand-sides of these equations
vanish we have that the matrices give the ADHM data for the
construction of charge $v$ $SU(w)$ instantons on $\R\sp4$. For
non-vanishing right-hand-side a modified ADHM constuction yields
instantons on a non-commutative space \cite{neksch}. One can also
apply the usual ADHM construction to the matrices above \cite{BN}.
The gauge fields resulting will not of course be \asd\ with respect
to the standard metric but one can ask whether they have any
further nice properties. We shall perform this construction and in
the process encounter several surprises. First we will see that
$U(1)$ instantons exist and are well behaved. Such instantons do
not exist on $\R\sp4$ but exist here because, as we shall
discover, space-time is ``blown-up". On this blown-up space we
will show there is a natural metric for which the charge one
abelian instanton is in fact self-dual. The higher charge case
will be dealt with elsewhere.

\section{An Integrable System}
The class of integrable systems we shall focus on here are of the
Calogero-Moser family \cite{Ca,cal,OP1}.
These systems have a rather rich structure
with connections to representation theory \cite{mat}, functional equations and
index theorems \cite{Ca2,BCa,BCb,BB1,BB2,bp,Gu,HBJ}
to Seiberg-Witten and topological field theory \SWrefs.
The quantum mechanics of these systems has been well studied
\cite{cal,OP2,fv,suth,Su2,OOS,OS,kt}.
Many properties of these models can be found in the book \cite{cmb}.
The Calogero-Moser
systems are in many ways generic: given an integrable system with polynomial
conserved quantities and suitable symmetry we arrive at these models. Thus
for example we may characterise the ($a_n$) Calogero-Moser system
by \cite{B}:
\begin{thm}
Let $H$ and $P$ be the (natural) Hamiltonian and centre of mass momentum
\begin{equation*}
H= \frac{1}{2} \sum\limits_{i=1}^{n}p_{i}^{2}+V,\qquad
P= \sum\limits_{i=1}^{n}p_{i}.
\label{HP}
\end{equation*}
Denote by $Q$ an independent third order quantity
\begin{equation*}
Q=\sum\limits_{i=1}^{n}p_{i}^{3}+
\frac{1}{6}\sum\limits_{i\ne j\ne k}d_{ijk}p_i p_j p_k
+\sum\limits_{i\ne j}d_{ij}p_i^2  p_j+
\frac{1}{2}\sum\limits_{ij}a_{ij}p_{i}p_{j}+\sum_ib_{i}p_{i} +c.
\label{Q}
\end{equation*}
If these are $S_{n}$ invariant and Poisson commute,
\begin{equation*}
\left\{ P,H\right\} =\left\{ P,Q\right\} =\left\{ Q,H\right\} =0,
\end{equation*}
then
$V=\frac{1}{6}\sum\limits_{i\neq j}\wp\left( x_{i}-x_{j}\right) +const$
and we have the Calogero-Moser system.
\end{thm}

For the our purposes here we will need only the simplest example. Following
Olshanetsky and Perelomov we will derive these models as a coadjoint
reduction of a simpler system \cite{OP1, kkk}.
Consider a Lie algebra $\mathfrak g$ and
$A\in\mathfrak{g}$ moving freely on this vector space: $\ddot A=0$. Thus
$A=a+bt$ for constant $a$, $b\in\mathfrak{g}$. We may conjugate $A$ to
give a piece $X_{ss}$ lying in a given Cartan subalgebra and a constant
(possibly vanishing) nilpotent piece $X_n$:
$$A=gXg\sp{-1},$$
with $X=X_{ss}+X_n$. For $gl(v)$ this is simply putting $A$ into Jordan form.
Now
\begin{align*}
\dot A&=g\left( \dot X +[M,X]\right)g\sp{-1}:=g Lg\sp{-1},
\qquad M=g\sp{-1}\dot g, \\
\ddot A&=g\left( \dot L +[M,L]\right)g\sp{-1}=0.
\end{align*}
Thus we obtain a Lax pair (without spectral parameter)
$\dot L=[L,M]$ of the form $L=\dot X +[M,X]$
corresponding to geodesics $\ddot A=0$. Now consider the obviously conserved
angular momentum
\begin{equation}\label{amdef}
[A,\dot A]=g[X,L]g\sp{-1}:=g\bar C g\sp{-1}
\end{equation}
where clearly
\begin{equation}\label{amev}
0=\dot{ \bar C} +[M,\bar C].
\end{equation}
When $X$ is semi-simple it is particularly easy to solve for $L$ in terms
of $\bar C$ and $X$. Considering the case of $gl(v)$ and assuming $X$
semi-simple, or equivalently that $A(t)$ is diagonalisable, we obtain
$$L= {\dot x}\cdot H +\sum_{\alpha\in\Phi}
\frac{\left(\Tr {\bar C}E_{-\alpha}\right)}{\alpha\cdot x}E_\alpha
=\sum_i \dot x_i H_i +\sum_{i\ne j}
\frac{\left(\Tr {\bar C}E_{ji}\right)}{x_i-x_j}E_{ij}.
$$
Here $\{H_i, E_\alpha\}$ form a Chevalley basis of $\mathfrak {g}$ with root
system $\Phi$ and we have used a normalization $\Tr E_\alpha E_{-\beta}=
\delta_{\alpha,\beta}$.
As we see from (\ref{amev}) the matrix $\bar C$ is in general time-dependent
and this leads to the spin Calogero-Moser models. For particular angular
momentum it is possible to have a simplification: $\bar C$ will be constant
if and only if $[M,{\bar C}]=0$. The usual $gl(w)$ (spinless) Calogero-Moser
model corresponds to
$${\bar C}={\zeta_\C}\sum_\alpha E_\alpha = u\sp{T} u-{\zeta_\C}{\bf 1}_V$$
where $u=\sqrt{\zeta_\C}(1,1,\ldots,1)$. This yields the Lagrangian
$$\frac{1}{2}\Tr {\dot A}\sp2=\frac{1}{2}\Tr L\sp2=
\frac{1}{2}\sum_i {\dot x_i}\sp2 -
 \sum_{i< j}\frac{\zeta_\C\sp2}{(x_i-x_j)\sp2}
$$
and Hamiltonian
\begin{equation}\label{cmdef}
H=\frac{1}{2}\sum_i {p_i}\sp2 +
 \sum_{i< j}\frac{\zeta_\C\sp2}{(x_i-x_j)\sp2}.
\end{equation}
This Hamiltonian is that of the completely integrable system we are
interested in. By setting
\begin{equation}\label{cmcor}
B_1=L,\qquad  B_2=X,\qquad  I=u\sp{T},\qquad  J=u
\end{equation}
(or equivalently $B_1=\dot A$, $ B_2=A$, $ I=gu\sp{T}$, $ J=u g\sp{-1}$)
we have (\ref{cmom}) and for the normal matrices we have assumed (for which
$g$ is unitary) then (\ref{rmom}) is identically satisfied. In this case
$w=1$ and clearly conjugation of $B_{1,2}$ by $GL(v)$ with an attendant
action on $I$ and $J$ does not effect the reduced system. As $B_{1,2}$
are determined by the initial conditions the equivalence classes of
solutions up to this action describe the phase space of the Calogero-Moser
system. By considering the spin Calogero-Moser models we get $w>1$.

At this stage we have associated the rational (complexified)
Calogero-Moser integrable system with the equations (\ref{cmom}) and
(\ref{rmom}). Before looking at the this space of matrices more closely
it is worth recording the connection with Seiberg-Witten theory.
The moduli space of four-dimensional $N=2$ SYM with adjoint matter
is described by the elliptic Calogero-Moser model,
\begin{equation}\label{ecmdef}
H=\frac{1}{2}\sum_i {p_i}\sp2 +\zeta_\C\sp2 \sum_{i< j}\wp(x_i-x_j).
\end{equation}
Here the potential is described by the Weierstrass $\wp$-function
which has degenerations $1/\sin\sp2 x$ and $1/x\sp2$. The free
$N=4$ theory corresponds to the limit $\zeta_\C\sp2\rightarrow 0$
and there is also a double scaling limit in which the resulting
potential is that of the periodic Toda chain describes the pure
$N=2$ SYM gauge theory \cite{Inoz,inoz}. The perturbative limit of
the $N=2$ theory with adjoint matter is described in terms of the
potential $1/\sin\sp2 x$ while the perturbative limit of the pure
$N=2$ gauge theory is given by the non-periodic Toda chain. The
rational degeneration we are considering also has a field
theoretic interpretation: it arises in the perturbative limit of
the $N=2$ theory with massive adjoint matter, reduced down to
three dimensions.

\section{Moduli Spaces and Moment Maps}
We shall now describe in more detail the space of solutions to (\ref{cmom})
and (\ref{rmom}).
Let $V$ and $W$ be hermitian complex vector spaces of
dimensions $v$ and $w$ respectively and call $\alpha=(v,w)$ the dimension
vector. Let
$B_{1}$ and $B_{2}$ be maps from $V$ to itself, let
$I$ be a map from $W$ to $V$ and finally
let $J$ be a map from $V$ to $W$. This data may be expressed by
the quiver or directed graph below:

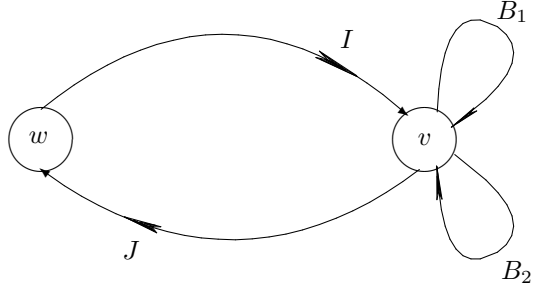
\special{em:linewidth 0.4pt} \unitlength 1.00mm
\linethickness{0.4pt}
\begin{picture}(96.00,89.00)
\put(29.00,64.67){\circle{8.00}}
\put(80.00,64.67){\circle{7.77}}
\put(77.67,68.00){\vector(4,-3){0.2}}
\emline{29.00}{68.67}{1}{31.22}{70.39}{2}
\emline{31.22}{70.39}{3}{33.43}{71.95}{4}
\emline{33.43}{71.95}{5}{35.63}{73.35}{6}
\emline{35.63}{73.35}{7}{37.83}{74.58}{8}
\emline{37.83}{74.58}{9}{40.03}{75.66}{10}
\emline{40.03}{75.66}{11}{42.22}{76.56}{12}
\emline{42.22}{76.56}{13}{44.40}{77.31}{14}
\emline{44.40}{77.31}{15}{46.58}{77.89}{16}
\emline{46.58}{77.89}{17}{48.75}{78.32}{18}
\emline{48.75}{78.32}{19}{50.92}{78.57}{20}
\emline{50.92}{78.57}{21}{53.08}{78.67}{22}
\emline{53.08}{78.67}{23}{55.24}{78.60}{24}
\emline{55.24}{78.60}{25}{57.40}{78.37}{26}
\emline{57.40}{78.37}{27}{59.54}{77.98}{28}
\emline{59.54}{77.98}{29}{61.69}{77.42}{30}
\emline{61.69}{77.42}{31}{63.82}{76.71}{32}
\emline{63.82}{76.71}{33}{65.95}{75.83}{34}
\emline{65.95}{75.83}{35}{68.08}{74.78}{36}
\emline{68.08}{74.78}{37}{70.20}{73.58}{38}
\emline{70.20}{73.58}{39}{72.32}{72.21}{40}
\emline{72.32}{72.21}{41}{74.43}{70.68}{42}
\emline{74.43}{70.68}{43}{77.67}{68.00}{44}
\put(29.00,60.67){\vector(-4,3){0.2}}
\emline{79.33}{60.67}{45}{77.12}{59.07}{46}
\emline{77.12}{59.07}{47}{74.91}{57.63}{48}
\emline{74.91}{57.63}{49}{72.70}{56.33}{50}
\emline{72.70}{56.33}{51}{70.48}{55.19}{52}
\emline{70.48}{55.19}{53}{68.25}{54.19}{54}
\emline{68.25}{54.19}{55}{66.02}{53.34}{56}
\emline{66.02}{53.34}{57}{63.79}{52.64}{58}
\emline{63.79}{52.64}{59}{61.56}{52.09}{60}
\emline{61.56}{52.09}{61}{59.32}{51.69}{62}
\emline{59.32}{51.69}{63}{57.07}{51.44}{64}
\emline{57.07}{51.44}{65}{54.82}{51.34}{66}
\emline{54.82}{51.34}{67}{52.57}{51.39}{68}
\emline{52.57}{51.39}{69}{50.32}{51.58}{70}
\emline{50.32}{51.58}{71}{48.06}{51.93}{72}
\emline{48.06}{51.93}{73}{45.79}{52.42}{74}
\emline{45.79}{52.42}{75}{43.52}{53.07}{76}
\emline{43.52}{53.07}{77}{41.25}{53.86}{78}
\emline{41.25}{53.86}{79}{38.97}{54.81}{80}
\emline{38.97}{54.81}{81}{36.69}{55.90}{82}
\emline{36.69}{55.90}{83}{34.41}{57.14}{84}
\emline{34.41}{57.14}{85}{32.12}{58.54}{86}
\emline{32.12}{58.54}{87}{29.00}{60.67}{88}
\emline{81.67}{68.33}{89}{81.88}{71.38}{90}
\emline{81.88}{71.38}{91}{82.23}{73.99}{92}
\emline{82.23}{73.99}{93}{82.73}{76.17}{94}
\emline{82.73}{76.17}{95}{83.38}{77.92}{96}
\emline{83.38}{77.92}{97}{84.17}{79.23}{98}
\emline{84.17}{79.23}{99}{85.10}{80.11}{100}
\emline{85.10}{80.11}{101}{86.18}{80.55}{102}
\emline{86.18}{80.55}{103}{87.41}{80.56}{104}
\emline{87.41}{80.56}{105}{89.67}{79.67}{106}
\emline{89.67}{79.67}{107}{90.87}{78.63}{108}
\emline{90.87}{78.63}{109}{91.60}{77.47}{110}
\emline{91.60}{77.47}{111}{91.86}{76.21}{112}
\emline{91.86}{76.21}{113}{91.64}{74.83}{114}
\emline{91.64}{74.83}{115}{90.95}{73.34}{116}
\emline{90.95}{73.34}{117}{89.78}{71.74}{118}
\emline{89.78}{71.74}{119}{88.14}{70.03}{120}
\emline{88.14}{70.03}{121}{84.00}{66.67}{122}
\emline{81.67}{61.00}{123}{81.88}{57.95}{124}
\emline{81.88}{57.95}{125}{82.23}{55.34}{126}
\emline{82.23}{55.34}{127}{82.73}{53.16}{128}
\emline{82.73}{53.16}{129}{83.38}{51.42}{130}
\emline{83.38}{51.42}{131}{84.17}{50.11}{132}
\emline{84.17}{50.11}{133}{85.10}{49.23}{134}
\emline{85.10}{49.23}{135}{86.18}{48.79}{136}
\emline{86.18}{48.79}{137}{87.41}{48.78}{138}
\emline{87.41}{48.78}{139}{89.67}{49.67}{140}
\emline{89.67}{49.67}{141}{90.87}{50.71}{142}
\emline{90.87}{50.71}{143}{91.60}{51.86}{144}
\emline{91.60}{51.86}{145}{91.86}{53.13}{146}
\emline{91.86}{53.13}{147}{91.64}{54.50}{148}
\emline{91.64}{54.50}{149}{90.95}{55.99}{150}
\emline{90.95}{55.99}{151}{89.78}{57.59}{152}
\emline{89.78}{57.59}{153}{88.14}{59.30}{154}
\emline{88.14}{59.30}{155}{84.00}{62.67}{156}
\emline{70.97}{73.16}{157}{66.00}{76.42}{158}
\emline{66.00}{76.42}{159}{67.10}{75.26}{160}
\emline{67.10}{75.26}{161}{65.55}{75.57}{162}
\emline{65.55}{75.57}{163}{71.07}{73.06}{164}
\emline{71.07}{73.06}{165}{66.15}{75.37}{166}
\emline{66.15}{75.37}{167}{69.56}{73.91}{168}
\emline{69.56}{73.91}{169}{66.85}{75.26}{170}
\emline{66.85}{75.26}{171}{69.16}{74.21}{172}
\emline{69.16}{74.21}{173}{66.35}{76.12}{174}
\emline{66.35}{76.12}{175}{68.21}{74.71}{176}
\emline{68.21}{74.71}{177}{67.25}{75.26}{178}
\emline{40.11}{54.39}{179}{44.94}{53.05}{180}
\emline{44.94}{53.05}{181}{43.70}{53.05}{182}
\emline{43.70}{53.05}{183}{44.40}{52.24}{184}
\emline{44.40}{52.24}{185}{40.11}{54.39}{186}
\emline{40.11}{54.39}{187}{44.13}{52.46}{188}
\emline{44.13}{52.46}{189}{41.34}{53.96}{190}
\emline{41.34}{53.96}{191}{43.86}{52.73}{192}
\emline{43.86}{52.73}{193}{42.15}{53.69}{194}
\emline{42.15}{53.69}{195}{44.24}{53.21}{196}
\emline{44.24}{53.21}{197}{42.79}{53.48}{198}
\emline{42.79}{53.48}{199}{43.97}{53.10}{200}
\emline{43.97}{53.10}{201}{43.33}{53.15}{202}
\put(28.67,65.00){\makebox(0,0)[cc]{$w$}}
\put(69.67,78.00){\makebox(0,0)[cc]{$I$}}
\put(41.00,50.00){\makebox(0,0)[cc]{$J$}}
\put(80.00,64.67){\makebox(0,0)[cc]{$v$}}
\put(91.67,81.67){\makebox(0,0)[cc]{$B_1$}}
\put(92.33,47.00){\makebox(0,0)[cc]{$B_2$}}
\emline{83.56}{66.32}{203}{86.92}{68.45}{204}
\emline{86.92}{68.45}{205}{85.98}{68.15}{206}
\emline{85.98}{68.15}{207}{86.53}{68.87}{208}
\emline{86.53}{68.87}{209}{83.60}{66.32}{210}
\emline{83.60}{66.32}{211}{86.28}{68.53}{212}
\emline{86.28}{68.53}{213}{84.45}{66.92}{214}
\emline{84.45}{66.92}{215}{86.45}{68.28}{216}
\emline{86.45}{68.28}{217}{85.39}{67.68}{218}
\emline{81.64}{61.00}{219}{81.64}{56.67}{220}
\emline{81.64}{56.67}{221}{81.85}{57.69}{222}
\emline{81.85}{57.69}{223}{82.32}{56.75}{224}
\emline{82.32}{56.75}{225}{81.60}{61.13}{226}
\emline{81.60}{61.13}{227}{82.15}{57.22}{228}
\emline{82.15}{57.22}{229}{81.68}{59.73}{230}
\emline{81.68}{59.73}{231}{81.73}{57.43}{232}
\emline{81.73}{57.43}{233}{81.77}{59.05}{234}
\emline{81.77}{59.05}{235}{81.90}{57.64}{236}
\end{picture}
\vskip-1.6in

The space of matrices $$V_{(v,w)}=\{(B_{1}, B_{2}, I, J)|\
B_{1,2}\in M_v(\C),\ I\in M_{v\times w}(\C), \ J\in M_{w\times
v}(\C)\}$$ appearing here is a (flat) hyper-K\"ahler manifold. We
have a metric on $V_{(v,w)}$ coming from the hermitian inner
product $<\alpha,\beta>=\Tr \alpha \beta\sp\dagger$ on $M_{r\times
s}(\C)$ matrices. With $x=(B_{1}, B_{2}, I, J)$ and $y=(\tilde
B_{1}, \tilde B_{2},\tilde I, \tilde J)$ this is given by
$$g(x,y)=\frac{1}{2}\Tr\left(B_{1}\tilde B_{1}\sp\dagger +\tilde
B_1 B_1\sp\dagger+ B_{2}\tilde B_{2}\sp\dagger +\tilde B_2
B_2\sp\dagger+ I\tilde I\sp\dagger+\tilde I I\sp\dagger
+J\sp\dagger \tilde J+ \tilde J\sp\dagger J\right). $$ This metric
is hermitian for the three complex structures
\begin{align}
\hat i \cdot (B_{1}, B_{2}, I, J)&=(iB_1,iB_2,iI,iJ),\\ \hat j
\cdot (B_{1}, B_{2}, I,J) &=
(-B_2\sp\dagger,B_1\sp\dagger,-J\sp\dagger,I\sp\dagger),\\ \hat
k&=\hat i\hat j,
\end{align}
which obey the usual relations of the quaternions. That is
$$g(x,y)=g(\hat i x,\hat i y)=g(\hat j x,\hat j y)=g(\hat k x,\hat k y),$$
and ${\hat i}\sp2={\hat j}\sp2={\hat k}\sp2=\hat i \hat j \hat k=-1$.
We have associated to each of the complex structures  the K\"ahler forms
\begin{align*}
\omega_1(x,y)&=g(\hat i x, y)=\frac{i}{2}\Tr\left( B_{1}\tilde
B_{1}\sp\dagger -\tilde B_1 B_1\sp\dagger+ B_{2}\tilde
B_{2}\sp\dagger -\tilde B_2 B_2\sp\dagger+ I\tilde
I\sp\dagger-\tilde I I\sp\dagger -J\sp\dagger \tilde J+\tilde
J\sp\dagger J \right),\\ \omega_2(x,y)&=g(\hat j x,
y)=-\frac{1}{2}\Tr\left( \tilde B_{1} B_2 +B_2\sp\dagger \tilde
B_1\sp\dagger-\tilde B_{2} B_1-
 B_1\sp\dagger \tilde B_2\sp\dagger+
\tilde I J+ J\sp\dagger \tilde I\sp\dagger - I\tilde J-
 \tilde J\sp\dagger I\sp\dagger\right),\\
\omega_3(x,y)&=g(\hat k x, y)=\frac{i}{2}\Tr\left( \tilde B_{1}
B_2 -B_2\sp\dagger \tilde B_1\sp\dagger-\tilde B_{2} B_1+
 B_1\sp\dagger \tilde B_2\sp\dagger+
\tilde I J- J\sp\dagger \tilde I\sp\dagger - I\tilde J+
 \tilde J\sp\dagger I\sp\dagger\right).
\end{align*}
We may express the first symplectic form as
\begin{align*}
\omega_1&=\frac{i}{2}\Tr\left( dB_1 \wedge dB_1\sp\dagger+
dB_2 \wedge dB_2\sp\dagger +d I\wedge dI\sp\dagger-dJ\sp\dagger\wedge dJ \right)\\
&=
\frac{i}{2} d\Tr\left(B_1  dB_1\sp\dagger+B_2 dB_2\sp\dagger +
IdI\sp\dagger-J\sp\dagger dJ\right)
=\d \Theta_1,
\end{align*}
which shows it closed. Similarly for the other K\"ahler forms we have
\begin{align*}
\omega_2&=\frac{1}{2}d\Tr\left(B_1  dB_2+B_1\sp\dagger
dB_2\sp\dagger + I dJ- J\sp\dagger d I\sp\dagger\right) =\d
\Theta_2,\\ \omega_3&=\frac{1}{2i} d\Tr\left(B_1
dB_2-B_1\sp\dagger dB_2\sp\dagger + I dJ+ J\sp\dagger d
I\sp\dagger\right) =\d \Theta_3.
\end{align*}
It is also convenient to introduce
$$\omega_\C(x,y)=\omega_2(x,y)+i\omega_3(x,y)= \Tr\left( B_{2}
\tilde B_1- B_1 \tilde B_2+
 J \tilde I-\tilde J  I \right).
$$ We see from $\omega_\C(\hat i u,v)=g(\hat j\hat i u,v)+i g(\hat
k\hat i u,v)=-\omega_3(u,v)+i\omega_2(u,v)=i\omega_\C(u,v)$ that
this is of type $(2,0)$. We can write this complex symplectic form
as $$\omega_\C=\Tr\left( dB_1 \wedge dB_2
 +d I \wedge d J \right)=
d\Tr\left(B_1  dB_2+I  d J \right)=\d \Theta_\C $$ with
$\Theta_\C=\Theta_2+i\Theta_3$.

There is also a natural action of $GL(\alpha):=GL(v,\C)\times GL(w,\C)$ on
$V_{(v,w)}$ via
\be
\label{scndr}(g,h):\left( B_{1}, B_{2}, I, J \right)
\mapsto \left( gB_{1}g\sp{-1}, \, g B_{2} g\sp{-1}, \, g Ih\sp{-1} ,  \,
h J g \sp{-1} \right).
\ee
For $(g,h)\in U_v(\C)\times U_w(\C)$ we have that
$\hat j\circ (g,h)=(g,h)\circ\hat j$ and consequently $U_v(\C)\times U_w(\C)$
preserves the quaternionic structure of $V_{(v,w)}$.
Observe that although the metric and $\omega_1$ are only $U_v(\C)\times U_w(\C)$
invariant the complex symplectic form $\omega_\C$ is in fact
$GL(v,\C)\times GL(w,\C)$ invariant.
Associated with the $U_v(\C)$ action we have the tangent vector
$x_\xi=([\xi,B_1],[\xi,B_2],\xi I, -J\xi)$ and we may associate a Hamiltonian
and moment map
to each of the symplectic structures via
$$H_i\sp\xi=\Theta_i(x_\xi)=<\xi\sp\dagger,\mu_i>.$$
Now we have
$$
\Theta_1(x_\xi)=\frac{i}{2}\Tr\xi\sp\dagger \left([B_1,B_1\sp\dagger]+
[B_2,B_2\sp\dagger]+II\sp\dagger-J\sp\dagger J\right)\sp\dagger=\frac{i}{2}
<\xi\sp\dagger,[B_1,B_1\sp\dagger]+
[B_2,B_2\sp\dagger]+II\sp\dagger-J\sp\dagger J>,
$$
and
\begin{align*}
\Theta_2(x_\xi)&=\frac{1}{2}
<\xi\sp\dagger,[B_1,B_2]-[B_1,B_2]\sp\dagger+IJ-J\sp\dagger I\sp\dagger >,\\
\Theta_3(x_\xi)&=-\frac{i}{2}
<\xi\sp\dagger,[B_1,B_2]+[B_1,B_2]\sp\dagger+IJ+J\sp\dagger I\sp\dagger >,
\end{align*}
(in which we  have used $\xi\sp\dagger=-\xi$). Further
$$
\Theta_\C(x_\xi)=\Tr\xi\sp\dagger \left([B_1,B_2]+IJ\right)\sp\dagger=
<\xi\sp\dagger,[B_1,B_2]+IJ>
$$
(which does not require $\xi\sp\dagger=-\xi$) and so $\mu_\C=\mu_2+i\mu_3$.
We thus have the moment map
$\boldsymbol{\mu}:V_{(v,w)}\rightarrow \R\sp3\otimes u_v(\C)$ given by
$(\mu_1,\mu_2,\mu_3)$ defined above. It will be convenient to set $\mu_\R=
i\mu_1$ and so we have moment maps
\begin{align}
\label{remu} 2\mu_R (B_1,B_2,I,J) & = [B_1, B_1\sp\dagger]+
[B_2,B_2\sp\dagger]+II\sp\dagger-J\sp\dagger J\\ \mu_\C
(B_1,B_2,I,J) & = [B_1,B_2] + IJ \ . \label{cmu}
\end{align}
We recognize (\ref{cmom}) and (\ref{rmom}) as fixing these moment
maps. Let us collect the numbers $({\z}_\R, {\rm Re} {\z}_{\C},
{\rm Im} {\z}_{\C})$ into a three-vector ${\vec{\z}}  \in {\R}^3$.
A standard result \cite{rkh} is that when $G=U_v(\C)$ acts freely
on $\mu\sp{-1}(\zeta)$ then $\mu\sp{-1}({\vec{\z}} )/G$ is a
smooth manifold with Riemannian metric and hyper-K\"ahler
structure induced from those on $V_{(v,w)}$.

Let us also record the moment map
$\boldsymbol{\tilde\mu}:V_{(v,w)}\rightarrow \R\sp3\otimes u_w(\C)$ for the
$U_w(\C)$ action given
$$\boldsymbol{\tilde\mu}(B_1,B_2,I,J)=\Big(\frac{i}{2}(JJ\sp\dagger-I\sp\dagger
I),\frac{1}{2}(I\sp\dagger J\sp\dagger-JI),\frac{i}{2}(I\sp\dagger
J\sp\dagger-JI)\Big).$$

Let us further consider the map $\mu_\C:V_{(v,w)}\rightarrow GL_v(\C)$ given
in (\ref{cmu}).
The differential $d\mu_\C$ may be determined from the the linear terms
in an $\epsilon$-expansion about $(B_1+\epsilon X,B_2+\epsilon Y,I+\epsilon
L,J+\epsilon M)$:
$$d\mu_c\big\vert_{(B_1,B_2,I,J)}(X,Y,L,M)=[X,B_1]+[B_2,Y]+IM+LJ.$$
The image of $d\mu_\C$ is orthogonal to those matrices $W$ such that
$$<d\mu_c\big\vert_{(B_1,B_2,I,J)}(X,Y,L,M),W>=0. $$
Using the nondegeneracy
and cyclicity of $\Tr$ this is equivalent to
$$\{W\in GL_v(\C)\,\vert\  [B_1,W\sp\dagger]=0,\ [B_2,W\sp\dagger]=0,\
W\sp\dagger I=0,\ J W\sp\dagger=0\}.$$
In particular the $w$ columns of $I$ are in $\Ker W$ and this space is
stable under left multiplication by $B_1$ and $B_2$.

We remark that one may associate with any a quiver a path algebra
$\M$ with generators corresponding to the vertices and directed
edges the associative algebra structure is induced by the
concatenation of paths where possible and zero otherwise. Thus to
the path algebra $\M$ associated with

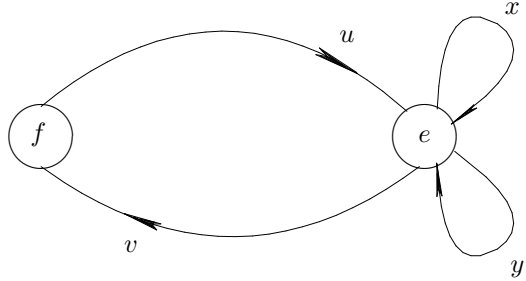
\special{em:linewidth 0.4pt} \unitlength 1.00mm
\linethickness{0.4pt}
\begin{picture}(96.00,89.00)
\put(29.00,64.67){\circle{8.00}}
\put(80.00,64.67){\circle{7.77}}
\emline{29.00}{68.67}{1}{31.22}{70.39}{2}
\emline{31.22}{70.39}{3}{33.43}{71.95}{4}
\emline{33.43}{71.95}{5}{35.63}{73.35}{6}
\emline{35.63}{73.35}{7}{37.83}{74.58}{8}
\emline{37.83}{74.58}{9}{40.03}{75.66}{10}
\emline{40.03}{75.66}{11}{42.22}{76.56}{12}
\emline{42.22}{76.56}{13}{44.40}{77.31}{14}
\emline{44.40}{77.31}{15}{46.58}{77.89}{16}
\emline{46.58}{77.89}{17}{48.75}{78.32}{18}
\emline{48.75}{78.32}{19}{50.92}{78.57}{20}
\emline{50.92}{78.57}{21}{53.08}{78.67}{22}
\emline{53.08}{78.67}{23}{55.24}{78.60}{24}
\emline{55.24}{78.60}{25}{57.40}{78.37}{26}
\emline{57.40}{78.37}{27}{59.54}{77.98}{28}
\emline{59.54}{77.98}{29}{61.69}{77.42}{30}
\emline{61.69}{77.42}{31}{63.82}{76.71}{32}
\emline{63.82}{76.71}{33}{65.95}{75.83}{34}
\emline{65.95}{75.83}{35}{68.08}{74.78}{36}
\emline{68.08}{74.78}{37}{70.20}{73.58}{38}
\emline{70.20}{73.58}{39}{72.32}{72.21}{40}
\emline{72.32}{72.21}{41}{74.43}{70.68}{42}
\emline{74.43}{70.68}{43}{77.67}{68.00}{44}
\emline{79.33}{60.67}{45}{77.12}{59.07}{46}
\emline{77.12}{59.07}{47}{74.91}{57.63}{48}
\emline{74.91}{57.63}{49}{72.70}{56.33}{50}
\emline{72.70}{56.33}{51}{70.48}{55.19}{52}
\emline{70.48}{55.19}{53}{68.25}{54.19}{54}
\emline{68.25}{54.19}{55}{66.02}{53.34}{56}
\emline{66.02}{53.34}{57}{63.79}{52.64}{58}
\emline{63.79}{52.64}{59}{61.56}{52.09}{60}
\emline{61.56}{52.09}{61}{59.32}{51.69}{62}
\emline{59.32}{51.69}{63}{57.07}{51.44}{64}
\emline{57.07}{51.44}{65}{54.82}{51.34}{66}
\emline{54.82}{51.34}{67}{52.57}{51.39}{68}
\emline{52.57}{51.39}{69}{50.32}{51.58}{70}
\emline{50.32}{51.58}{71}{48.06}{51.93}{72}
\emline{48.06}{51.93}{73}{45.79}{52.42}{74}
\emline{45.79}{52.42}{75}{43.52}{53.07}{76}
\emline{43.52}{53.07}{77}{41.25}{53.86}{78}
\emline{41.25}{53.86}{79}{38.97}{54.81}{80}
\emline{38.97}{54.81}{81}{36.69}{55.90}{82}
\emline{36.69}{55.90}{83}{34.41}{57.14}{84}
\emline{34.41}{57.14}{85}{32.12}{58.54}{86}
\emline{32.12}{58.54}{87}{29.00}{60.67}{88}
\emline{81.67}{68.33}{89}{81.88}{71.38}{90}
\emline{81.88}{71.38}{91}{82.23}{73.99}{92}
\emline{82.23}{73.99}{93}{82.73}{76.17}{94}
\emline{82.73}{76.17}{95}{83.38}{77.92}{96}
\emline{83.38}{77.92}{97}{84.17}{79.23}{98}
\emline{84.17}{79.23}{99}{85.10}{80.11}{100}
\emline{85.10}{80.11}{101}{86.18}{80.55}{102}
\emline{86.18}{80.55}{103}{87.41}{80.56}{104}
\emline{87.41}{80.56}{105}{89.67}{79.67}{106}
\emline{89.67}{79.67}{107}{90.87}{78.63}{108}
\emline{90.87}{78.63}{109}{91.60}{77.47}{110}
\emline{91.60}{77.47}{111}{91.86}{76.21}{112}
\emline{91.86}{76.21}{113}{91.64}{74.83}{114}
\emline{91.64}{74.83}{115}{90.95}{73.34}{116}
\emline{90.95}{73.34}{117}{89.78}{71.74}{118}
\emline{89.78}{71.74}{119}{88.14}{70.03}{120}
\emline{88.14}{70.03}{121}{84.00}{66.67}{122}
\emline{81.67}{61.00}{123}{81.88}{57.95}{124}
\emline{81.88}{57.95}{125}{82.23}{55.34}{126}
\emline{82.23}{55.34}{127}{82.73}{53.16}{128}
\emline{82.73}{53.16}{129}{83.38}{51.42}{130}
\emline{83.38}{51.42}{131}{84.17}{50.11}{132}
\emline{84.17}{50.11}{133}{85.10}{49.23}{134}
\emline{85.10}{49.23}{135}{86.18}{48.79}{136}
\emline{86.18}{48.79}{137}{87.41}{48.78}{138}
\emline{87.41}{48.78}{139}{89.67}{49.67}{140}
\emline{89.67}{49.67}{141}{90.87}{50.71}{142}
\emline{90.87}{50.71}{143}{91.60}{51.86}{144}
\emline{91.60}{51.86}{145}{91.86}{53.13}{146}
\emline{91.86}{53.13}{147}{91.64}{54.50}{148}
\emline{91.64}{54.50}{149}{90.95}{55.99}{150}
\emline{90.95}{55.99}{151}{89.78}{57.59}{152}
\emline{89.78}{57.59}{153}{88.14}{59.30}{154}
\emline{88.14}{59.30}{155}{84.00}{62.67}{156}
\emline{70.97}{73.16}{157}{66.00}{76.42}{158}
\emline{66.00}{76.42}{159}{67.10}{75.26}{160}
\emline{67.10}{75.26}{161}{65.55}{75.57}{162}
\emline{65.55}{75.57}{163}{71.07}{73.06}{164}
\emline{71.07}{73.06}{165}{66.15}{75.37}{166}
\emline{66.15}{75.37}{167}{69.56}{73.91}{168}
\emline{69.56}{73.91}{169}{66.85}{75.26}{170}
\emline{66.85}{75.26}{171}{69.16}{74.21}{172}
\emline{69.16}{74.21}{173}{66.35}{76.12}{174}
\emline{66.35}{76.12}{175}{68.21}{74.71}{176}
\emline{68.21}{74.71}{177}{67.25}{75.26}{178}
\emline{40.11}{54.39}{179}{44.94}{53.05}{180}
\emline{44.94}{53.05}{181}{43.70}{53.05}{182}
\emline{43.70}{53.05}{183}{44.40}{52.24}{184}
\emline{44.40}{52.24}{185}{40.11}{54.39}{186}
\emline{40.11}{54.39}{187}{44.13}{52.46}{188}
\emline{44.13}{52.46}{189}{41.34}{53.96}{190}
\emline{41.34}{53.96}{191}{43.86}{52.73}{192}
\emline{43.86}{52.73}{193}{42.15}{53.69}{194}
\emline{42.15}{53.69}{195}{44.24}{53.21}{196}
\emline{44.24}{53.21}{197}{42.79}{53.48}{198}
\emline{42.79}{53.48}{199}{43.97}{53.10}{200}
\emline{43.97}{53.10}{201}{43.33}{53.15}{202}
\put(28.67,65.00){\makebox(0,0)[cc]{$f$}}
\put(69.67,78.00){\makebox(0,0)[cc]{$u$}}
\put(41.00,50.00){\makebox(0,0)[cc]{$v$}}
\put(80.00,64.67){\makebox(0,0)[cc]{$e$}}
\put(91.67,81.67){\makebox(0,0)[cc]{$x$}}
\put(92.33,47.00){\makebox(0,0)[cc]{$y$}}
\emline{83.56}{66.32}{203}{86.92}{68.45}{204}
\emline{86.92}{68.45}{205}{85.98}{68.15}{206}
\emline{85.98}{68.15}{207}{86.53}{68.87}{208}
\emline{86.53}{68.87}{209}{83.60}{66.32}{210}
\emline{83.60}{66.32}{211}{86.28}{68.53}{212}
\emline{86.28}{68.53}{213}{84.45}{66.92}{214}
\emline{84.45}{66.92}{215}{86.45}{68.28}{216}
\emline{86.45}{68.28}{217}{85.39}{67.68}{218}
\emline{81.64}{61.00}{219}{81.64}{56.67}{220}
\emline{81.64}{56.67}{221}{81.85}{57.69}{222}
\emline{81.85}{57.69}{223}{82.32}{56.75}{224}
\emline{82.32}{56.75}{225}{81.60}{61.13}{226}
\emline{81.60}{61.13}{227}{82.15}{57.22}{228}
\emline{82.15}{57.22}{229}{81.68}{59.73}{230}
\emline{81.68}{59.73}{231}{81.73}{57.43}{232}
\emline{81.73}{57.43}{233}{81.77}{59.05}{234}
\emline{81.77}{59.05}{235}{81.90}{57.64}{236}
\end{picture}
\vskip-1.6in

\noindent one has relations $$\begin{array}{llll}
e^2=e&f^2=f&e+f=1\\ e.x=x&e.y=y&e.u=u&f.v=v\\
x.e=x&y.e=e&v.e=v&u.f=f
\end{array}
$$
with all other products vanishing.
By a representation $\rep_\alpha(\M)$ of a quiver one means the assignment to
each vertex
a vector space and to each directed edge a linear map of the corresponding
vector spaces. The dimension-vector $\alpha$ of a representation is the integral
vector containing the dimensions of the vertex spaces. Representations
are defined up to equivalence of a change of basis in the vertex spaces.
We have thus been describing the representations a particular quiver.

\section{The Deformed Instanton Equations}
We shall now describe perhaps the simplest deformation of the instanton
equations. Form a sequence of linear maps
\be\label{monad}
V\ \ \mapright{ \sigma_z} \ \
V \otimes {\C}^2 \oplus W\ \ \mapright{ \tau_z} \ \ V
\ee
where
\be
\label{mps}
{\s_z} =
\left(\begin{array}{c}-B_{1} + z_1 \\ B_{2}- z_2 \\ J\end{array}\right),
\qquad {\tau}_z =\left(\begin{array}{ccc}B_{2}  - z_2 & B_{1} - z_1 & I
\end{array}\right).
\ee
Here $z=(z_1,z_2)\in \C\sp2$ are complex parameters and we let
$\s=\s_{(0,0)}$ etc.

Suppose now that the matrices $(B_{1,2}, I, J)$ obey the following equations:
\begin{align}
\label{mmnts1}
\tau_z \s_z \ & = \ {\zeta}_{\C} {\bf 1}_{V}, \\
\label{mmnts2}
\tau_z \tau_z^{\dagger} \ & = \ \Delta_z + \zeta_{\R} {\bf 1}_{V}, \\
\label{mmnts3}
{\s_z}^{\dagger}{\s_z} \ & =  \ \Delta_z - \zeta_{\R} {\bf 1}_{V}.
\end{align}
These may be rewritten as
\begin{align*}
[B_1,B_2]+IJ &= \ {\zeta}_{\C} {\bf 1}_{V},
\\
[B_1,B_1^{\dagger}]+[B_2,B_2^{\dagger}]+I I\sp\dagger-J\sp\dagger J &=
2\zeta_{\R} {\bf 1}_{V},
\end{align*}
which are precisely our (\ref{cmom}) and (\ref{rmom}), and here
\begin{align*}
\Delta_z &= I I\sp\dagger+J\sp\dagger J +\\&  \frac{1}{2}\left[ (
B_1 - z_1) ( B_1\sp\dagger - {\zb}_1 ) +( B_1\sp\dagger  -
{\zb}_1)(B_1 -z_1) +
 ( B_2 - z_2) ( B_2\sp\dagger - {\zb}_2) +(B_2\sp\dagger -
{\zb}_2)(B_2 - z_2)\right] \ .
\end{align*}
We observe a consequence of these equations is that
\be
[B_2-B_1\sp\dagger, B_2\sp\dagger +B_1]+(I+J\sp\dagger)(I\sp\dagger- J)=
(2\zeta_\R-\zeta_{\C}+\bar\zeta_{\C}){\bf 1}_{V}.
\ee
(From this we can deduce (\ref{rmom}) but not (\ref{cmom}).)
The previous section told us that the space of all matrices
$(B_{1}, B_{2}, I, J)$ is a hyper-K\"ahler vector space and the equations
(\ref{mmnts1}-\ref{mmnts3})
may be interpreted as $U_v(\C)$ hyper-K\"ahler moment maps \cite{rkh}.
We will denote by $\overline{M}_{v,w}=\mu\sp{-1}({\vec{\z}} )/U_v(\C)$,
the space of solutions to the equations (\ref{mmnts1}-\ref{mmnts3})
up to such a symmetry transformation.

When ${\vec{\z}}=0$ these equations, together with the injectivity and
surjectivity of ${\s_z}$ and ${\tau}_z$ respectively, yield the standard
ADHM construction. If one relaxes the injectivity condition
then one gets the Donaldson compactification of the instanton moduli
space \cite{donaldson}. In the nomenclature of Corrigan and Goddard
\cite{cg} describing charge $v$ $SU(w)$ instantons,
$${\boldsymbol \Delta}= \begin{pmatrix}-B_{1}& B_{2}\sp\dagger\\ B_{2} &
       B_{1}\sp\dagger\\ J&I\sp\dagger\end{pmatrix},$$
and
${\boldsymbol \Delta}\sp\dagger{\boldsymbol \Delta}=\Delta\otimes {\bf 1}_2$
corresponds to the equations (\ref{mmnts1}-\ref{mmnts3})  when ${\vec{\z}}=0$.
We are considering a deformation of the standard ADHM equations.
The moduli space $\overline{M}_{v,w}$ is the space of freckled instantons on
${\R}^4$ in the sense  of
\cite{freck}, a ``freckle" simply being a point at which $\sigma_z$ fails to be
injective.
One learns from \cite{nakajima} that the deformed ADHM data
parameterise the (semistable) torsion free sheaves on $\CP2$
whose restriction on the projective line ${\ell}_{\infty}$
at infinity is trivial.
Each torsion free sheaf ${\CE}$ is included into the exact sequence of sheaves
\begin{equation}
\label{exct}0 \longrightarrow {\CE} \longrightarrow
{\CF} \longrightarrow {\CS}_{Z} \longrightarrow 0
\end{equation}
where ${\CF}$ is a holomorphic bundle ${\CE}^{**}$ and ${\CS}_Z$ is a
skyscraper sheaf supported at points, the set $Z$ of freckles \cite{freck}.
{}From this exact sequence one learns that
\be
\label{chrns}
{\rm ch}_{i}({\CE}) = {\rm ch}_{i}({\CF}) - \# Z \delta_{i,2}.
\ee

The same equations (\ref{mmnts1}-\ref{mmnts3}) also describe a
further moduli space, those of instantons over a noncommutative
$\R\sp4$. In this work we have considered $z_{1,2}$ as ordinary
complex numbers and $\vec\zeta\ne0$. The same equations arise
however by considering $\vec\zeta=0$ and the space-time
coordinates having the following commutation relations:
$$[z_1,z_2]=-\z_\C,\qquad [z_1,\bar z_1 ]+[z_2,\bar z_2 ]=-2\z_\R.$$
An analogous construction to the ordinary (commutative) ADHM
construction produces noncommutative instantons \cite{neksch}.

Observe that by performing an $SU(2)$ transformation
\be
\label{suro}
\begin{pmatrix}B_{1} \\ B_{2}\end{pmatrix} \mapsto
\begin{pmatrix} {\a} B_{1} - {\b} B_{2}^{\dagger} \\
{\bar\a} B_{2} + {\bar \b} B_{1}^{\dagger} \end{pmatrix},\qquad
\begin{pmatrix}I \\ J\end{pmatrix}
\mapsto \begin{pmatrix} {\a} I - {\b} J^{\dagger} \\
{\bar\a} J + {\bar \b} I^{\dagger} \end{pmatrix},
\ee
with $\vert \a \vert^2 + \vert \b \vert^2 =1$, we can always rotate
$\vec\zeta$ into a vector $({\z}_\R , 0, 0)$. Such a transformation corresponds
to singling out a particular complex structure on our data, for which
$z=(z_1,z_2)$ are the holomorphic coordinates on the Euclidean
space-time. Further we may choose the complex structure such that
${\z}_\R>0$. For ${\z}_\R>0$ (\ref{mmnts2}) shows that
${\tau}_{z}{\tau}_{z}^{\dagger}$ is invertible and
$\tau_z$ is surjective. When ${\vec\zeta}=({\z}_\R , 0, 0)$ and $w=1$
\cite{nakajima} shows one can simplify the equations and set $J = 0$. Then
$I^{\dagger}I = 2v\zeta_\R$ and $[B_{1},B_{2}]=0$. This situation yields a
well-known moduli space, the Hilbert scheme of points on the plane which
we now recall.

\section{Hilbert Schemes of Points on the Plane}
A Hilbert scheme $X\sp{[n]}$ or $\Hilb_n(X)$
for a (suitable) manifold $X$ is roughly
speaking the moduli space of $n$ points on $X$. It differs in general from
the symmetric product $S\sp{n}X$ in that $X\sp{[n]}$ contains information
on how points collide: ``fat" points are points that have collided, and
the Hilbert scheme retains the directions in which they coalesced.
We have the (Hilbert-Chow) map $\pi: X\sp{[n]}\rightarrow S\sp{n}X$.
When $\dim X=1$ we have $X\sp{[n]}= S\sp{n}X$ because there is only one
direction for the points to collide. Thus for example the Hilbert scheme
of $n$ points in the affine line $\A$ (over an algebraically closed field $k$)
is
\begin{align*}
\A\sp{[n]}&=\{\CI\subset k[z]\,\vert \, \CI \ {\rm an\ ideal },\
 \dim_k k[z]/\CI=n\}\\
&=\{f(z)\in k[z]\,\vert\, f(z)=z^n+a_1 z^{n-1}+\ldots+a_n,\ a_i\in k\}\\
&=S\sp{n}\A.
\end{align*}
When $k=\C$ the Hilbert scheme often inherits nice properties
possessed by the base space $X$. Thus if $X$ has a holomorphic
symplectic form so does $X\sp{[n]}$ ($n\ge2$). When $X$ is a $K3$
or an abelian surface then $X\sp{[n]}$ has a hyper-K\"ahler
metric. These and many other results are described in
\cite{nakajima}.

For our purposes we will focus on the case when $\dim X=2$.
Various connections have been made between integrable systems and Hilbert schemes
in this dimension \cite{FGNR, GNR, takasaki}. Consider
the case when $X=\A\sp2$, the affine plane (over an algebraically closed
field $k$). Then
$$(\A\sp2)\sp{[n]}=\{\CI\subset k[z_1,z_2]\,\vert \, \CI \ {\rm an\ ideal },\
 \dim_k k[z_1,z_2]/\CI=n\}
$$
What is particularly convenient for us is the alternative description
\begin{thm}
There exists an isomorphism
$$(\A\sp2)\sp{[n]}\cong\frac{\left\{(B_1,B_2,I)\,\Big\vert\,
\begin{array}{rl} (i)& [B_1,B_2]=0,\\
(ii)&\, \nexists\ no\ subspace\ S\subsetneqq k\sp{n}\ such\ that\
B_{1,2}S\subset S\ and\ \Image\ I\subset S
\end{array}\right\} }{GL(n,k)}
$$
where $B_{1,2}\in \End(k\sp{n})$ and $I\in \Hom(k,k\sp{n})$ with the
$GL(n,k)$ action given by
$$g\cdot(B_1,B_2,I)=(gB_1 g\sp{-1},gB_2 g\sp{-1}, gI).$$
\end{thm}
The proof is constructive. Let $\CI$ be an ideal in $k[z_1,z_2]$ and define
$V=k[z_1,z_2]/\CI$. Let $B_{1,2}\in \End(V)$ be multiplication by $z_{1,2}$
mod $\CI$. Then $[B_1,B_2]=0$. Define $I\in \Hom(k,V)$ by $I(1)\equiv1$
mod $\CI$. Since $1$ multiplied by products of $z_1$ and $z_2$ spans all of
$k[z_1,z_2]$ then $k[B_1,B_2]I(1)=k\sp{n}$ and stability follows.

Conversely, given $(B_1,B_2,I)$ as in the theorem define
$\phi:k[z_1,z_2]\twoheadrightarrow k\sp{n}$ by $\phi\left(
f(z_1,z_2)\right)=f(B_1,B_2)I(1)$. This is well defined by (i) and since
$\Image \phi$ is $B_{1,2}$ invariant and contains $\Image I$ then by
the stability (ii) it must be all of $k\sp n$. Thus $\phi$ is onto and
$\Ker\phi$ is then a codimension $n$ ideal $\CI\lhd k[z_1,z_2]$ yielding
a point of $(\A\sp2)\sp{[n]}$.  Thus
$$\CI=\{f(z_1,z_2)\in k[z_1,z_2]\, |\,f(B_1,B_2)I(1)=0\}
=\{f(z_1,z_2)\in k[z_1,z_2]\, |\,f(B_1,B_2)=0\}
$$
where the last equality follows from the stability condition.

\noindent{\bf Example $\Hilb_1(\C\sp2)$:} In this case $\dim_\C V=1$ and
so $B_1=\lambda$, $B_2=\mu$ are scalars. The stability condition means we
require $I=I(1)\ne0$ and using the $GL(1,\C)=\C\sp{\star}$ invariance
we may scale so that $I=1$. Thus $\Hilb_1(\C\sp2)\cong \{(\lambda,\mu,1)\in
\C\sp3\}\cong\C\sp2$. The ideal $\CI$ in this case is
$$\CI=\{f(z_1,z_2)\in k[z_1,z_2]\,
|\,f(\lambda,\mu)=0\}=<z_1-\lambda,z_2-\mu>,$$
that is the maximal $\mathfrak{m}_p$ ideal corresponding to the point
$p=(\lambda,\mu)\in\C\sp2$.

\noindent{\bf Example $\Hilb_2(\C\sp2)$:}
Now $\dim_\C V=2$ and $B_{1,2}$ are $2\times2$ matrices. We will consider
two cases: $B_1$ diagonalisable with distinct eigenvalues and $B_1$
not diagonalisable.

First suppose
$B_1=\left(\begin{array}{cc}\lambda_1&0\\0&\lambda_2\end{array}\right)$
where $\lambda_1\ne\lambda_2$. The commutativity of $B_1$ and $B_2$ yields
$B_2=\left(\begin{array}{cc}\mu_1&0\\0&\mu_2\end{array}\right)$ where
we do not demand the distinctness of $\mu_1$, $\mu_2$.
Stability now yields that
$I=\left(\begin{array}{c}\nu_1\\ \nu_2\end{array}\right)$ where $\nu_1
\nu_2\ne0$, and using the group conjugation we may scale this so that
$I=\left(\begin{array}{c}1\\ 1\end{array}\right)$. Thus we have a
representative of this orbit as
$$(\left(\begin{array}{cc}\lambda_1&0\\0&\lambda_2\end{array}\right),
\left(\begin{array}{cc}\mu_1&0\\0&\mu_2\end{array}\right),
\left(\begin{array}{c}1\\ 1\end{array}\right)).$$
The ideal corresponding to this is
$$
\CI=\{f(z_1,z_2)\in k[z_1,z_2]\,
|\,f(\lambda_1,\mu_1)=0=f(\lambda_2,\mu_2)\},$$
or $\mathfrak{ m}_{p_1}\cap \mathfrak{ m}_{p_2}$
which represents the two distinct points $p_1=(\lambda_1,\mu_1)$
and $p_2=(\lambda_2,\mu_2)$ in $\C\sp2$.

Next consider the situation when $B_1$ is not diagonalisable. Then $B_1$
can be taken to have Jordan form
$B_1=\left(\begin{array}{cc}\lambda&1\\0&\lambda\end{array}\right)$
and $B_2$ is found to be
$\left(\begin{array}{cc}\mu&\star\\0&\mu\end{array}\right)$. Similarly
we find $I=\left(\begin{array}{c}\nu_1\\ \nu_2\end{array}\right)$ where
$\nu_2\ne0$. A representative for this orbit can then be taken to be
\begin{equation}
(\left(\begin{array}{cc}\lambda&\alpha\\0&\lambda\end{array}\right),
\left(\begin{array}{cc}\mu&\beta\\0&\mu\end{array}\right),
\left(\begin{array}{c}0\\ 1\end{array}\right))\qquad{\rm where}\qquad
[\alpha:\beta\,]\in \CP1.
\label{hilb2e}
\end{equation}
It remains to describe the ideal associated with this orbit type. Using
$$\left(\begin{array}{cc}\lambda&\alpha\\0&\lambda\end{array}\right)\sp{k}
\left(\begin{array}{cc}\mu&\beta\\0&\mu\end{array}\right)\sp{l}=
\left(\begin{array}{cc}
\lambda\sp{k}\mu\sp{l}&
k\alpha\lambda\sp{k-1}\mu\sp{l}+l\beta\lambda\sp{k}\mu\sp{l-1}\\
0&\lambda\sp{k}\mu\sp{l} \end{array}\right).
$$
we find that we can represent $\CI$ as
\begin{align*}
\CI&=\{f(z_1,z_2)\in k[z_1,z_2]\, |\,f(\lambda,\mu)=0=
\alpha\partial_{z_1}f(z_1,z_2)\big\vert_{(\lambda,\mu)}+
\beta\partial_{z_2}f(z_1,z_2)\big\vert_{(\lambda,\mu)}\}\\
&=
<(z_1-\lambda)\sp2, (z_1-\lambda)(z_2-\mu),(z_2-\mu)\sp2,
\beta(z_1-\lambda)-\alpha(z_2-\mu)>.
\end{align*}
We can picture this as two points which have coalesced to the point
$p=(\lambda,\mu)$ colliding with
each other in the direction $\alpha\partial_{z_1}+
\beta\partial_{z_2}$. For each point in $\C\sp2$ there is a family
$[\alpha:\beta\,]\in \CP1$ of such fat points.

Now the two cases just given in fact exhaust the possible orbit types of
$\Hilb_2(\C\sp2)$ up to the interchange of $B_1$ and $B_2$.
If $B_1$ is in fact diagonal with equal eigenvalues then $B_2$ may be
diagonal with distinct eigenvalues which is the first case above, or it may
be nondiagonalisable and so in the second case. The only remaining
possibility is that both $B_1$ and $B_2$ are scalar multiples of the
identity, but this situation is ruled out by the stability requirement
as here $\C[B_1,B_2]$ gives a one-dimensional subspace of $\C\sp2$.

The Hilbert-Chow map $\pi:\Hilb_2(\C\sp2)\rightarrow S\sp2\C\sp2$ in this example
gives $\pi(B_1,B_2,I)=[p_1]+[p_2]$ for the
first case and $2[p]$ for the second case. Away from the diagonal we have a
one-to-one correspondence while on the diagonal the fibers are $\CP1$. In
fact $S\sp2\C\sp2$ has singularities and $\Hilb_2(\C\sp2)$ is smooth and gives
a resolution of these singularities.

\section{Constructing the gauge field}
We shall now construct a gauge field corresponding to the deformed
instanton equations following \cite{BN}. Our purpose is to further
investigate the properties of these gauge fields. The fundamental
object in the ADHM construction is the solution of
\begin{equation}
\label{ferm}
{\CD}^{\dagger}_{z} \Psi_{z} = 0, \qquad {\Psi}_{z} : W \to
V \otimes {\C}^2 \oplus W
\end{equation}
where
$${\CD}^{\dagger}_{z} = \begin{pmatrix}{\tau}_{z} \\ {\s}^{\dagger}_{z}
\end{pmatrix}.$$
We shall need the components
\begin{equation}\label{cmps}
\Psi_{z} = \begin{pmatrix}\Psi_{1} \\ \Psi_{2} \\ \chi\end{pmatrix}
= \begin{pmatrix}\vf \\ \chi\end{pmatrix}, \quad
\Psi_{1,2} \in V,\,\, {\vf} \in V \otimes{\C}^2 ,\,\,
{\chi} \in W.
\end{equation}

The solution of (\ref{ferm}) is not uniquely defined and one is free
to perform a $GL(w, {\C})$ gauge transformation,
$$\Psi_{z} \to \Psi_{z} \, g(z, {\zb}), \quad
  g(z, {\zb}) \in {\rm GL}(w, {\C}). $$
This gauge freedom can be partially fixed
by normalising the vector $\Psi_{z}$ as follows:
\begin{equation}\label{nrm}
\Psi_{z}^{\dagger} \Psi_{z} = {\bf 1}_{W}.
\end{equation}
With this normalisation the $U(w)$ gauge field is given by
\begin{equation}\label{gf}
A = {\Psi}^{\dagger}_{z} {\rm d} {\Psi}_{z},
\end{equation}
and its curvature is given by
\begin{equation}\label{vt}
F = {\Psi}^{\dagger}_{z} {\rm d}{\CD}_{z}
\frac{1}{ {\CD}^{\dagger}_{z} {\CD}_{z}}
{\rm d}{\CD}^{\dagger}_{z} {\Psi}_{z}.
\end{equation}

More explicitly,
$${\CD}^{\dagger}_z {\CD}_z = \Delta_z \otimes {\bf 1} -
{\bf 1}_{V} \otimes \zeta^{a} {\s}_{a},$$ hence
$$
\frac{1}{ {\CD}^{\dagger}_z {\CD}_z} =\frac{1}{{\Delta}^2_z - \vec\zeta^2}
\left( \Delta_z \otimes {\bf 1} + {\bf 1}_{V} \otimes {\zeta}^{a}{\s}_{a}
\right).
$$
Formula (\ref{vt}) makes sense for $z \in X^{\circ}\equiv{\R}^4 \setminus Z$,
where $X^{\circ}$ is the complement in ${\R}^4$ to the set $Z$ of points
(freckles) at which
\begin{equation}\label{dis}
{{\rm Det} \left( \Delta^2_z -  \vec\zeta^2 \right)= 0. }
\end{equation}
More explicitly (\ref{vt}) is
\begin{align}
F&=
{\vf}^{\dagger}
     \frac{\Delta_z \otimes \sigma_3}{{\Delta}^2_{z} - {\vec \zeta}^2}
{\vf}(d\bar z_1\wedge dz_1 -d\bar z_2\wedge dz_2 )
+2\, {\vf}^{\dagger}
     \frac{\Delta_z \otimes \sigma_+}{{\Delta}^2_{z} - {\vec \zeta}^2}
{\vf}\, d\bar z_1\wedge dz_2 \nonumber \\
&\qquad +2\, {\vf}^{\dagger}
     \frac{\Delta_z \otimes \sigma_-}{{\Delta}^2_{z} - {\vec \zeta}^2}
{\vf}\, d\bar z_2\wedge dz_1
+2i\,{\vf}^{\dagger} \frac{1}{{\Delta}^2_{z} - {\vec \zeta}^2} {\vf}
\,  \hat \zeta
\label{vtex}
\end{align}
where
$\hat\z = \z_{\R} {\varpi}_{\R} + {\z}_{\C} {\bar\varpi}_{\C}+
{\bar\z}_{\C}{\varpi}_{\C}$, and
\begin{equation}\label{sd}
\varpi_{\R} = \frac{i}{ 2}
\left( {\dd}z_1 \wedge    {\dd}\zb_1 +
{\dd}z_2 \wedge {\dd}\zb_2 \right), \,
\varpi_{\C} = {\dd}z_1 \wedge {\dd}z_2.
\end{equation}
With respect to the orientation given by $\star1=-dx^0\wedge
dx^1\wedge dx^2\wedge dx^3=-dz_1\wedge d\bar z_1\wedge dz_2\wedge
d\bar z_2$ where $z_1=(x^1+ix^2)/\sqrt{2}$ and
$z_2=(x^3+ix^0)/\sqrt{2}$ we have a basis of \asd 2-forms given by
\begin{align*}
\lambda_-\sp{1}&=dx\sp0\wedge dx\sp1+ dx\sp2\wedge dx\sp3= i(d\bar
z_2\wedge d z_1-dz_2\wedge d\bar z_1),\\
\lambda_-\sp{2}&=dx\sp0\wedge dx\sp2+ dx\sp3\wedge dx\sp1= \quad
d\bar z_2\wedge d z_1+dz_2\wedge d\bar z_1,\\
\lambda_-\sp{3}&=dx\sp0\wedge dx\sp3+ dx\sp1\wedge dx\sp2=
i(dz_1\wedge d\bar z_1-dz_2\wedge d\bar z_2),
\end{align*}
and a basis for the self-dual 2-forms given by
\begin{align*}
\lambda_+\sp{1}&=dx\sp0\wedge dx\sp1- dx\sp2\wedge dx\sp3= i( d
z_1\wedge dz_2- d\bar z_1\wedge d\bar z_2),\\
\lambda_+\sp{2}&=dx\sp0\wedge dx\sp2- dx\sp3\wedge dx\sp1= \quad d
z_1\wedge dz_2+ d\bar z_1\wedge d\bar z_2,\\
\lambda_+\sp{3}&=dx\sp0\wedge dx\sp3- dx\sp1\wedge dx\sp2=
i(dz_1\wedge d\bar z_1 +dz_2\wedge d\bar z_2).
\end{align*}
With this orientation we see $F$ is \asd when $\vec\zeta=0$. One
then has on $X^{\circ}$ that
\begin{equation}\label{sdp}
F^{+} :=\frac{1}{2}\left(F + \phantom{.}\sp\star F\right)= 4 i\,
{\vf}^{\dagger} \frac{1}{{\Delta}^2_{z} - {\vec \zeta}^2} {\vf} \,
\hat \zeta.
\end{equation}
Thus with respect to the standard complex coordinates the gauge field we
have constructed is neither self-dual nor anti-self-dual. We can ask whether
it has other nice properties. For example, do other coordinates exist for
which it is either self-dual or anti-self-dual?

If $\zeta^{\C}=0$ then (\ref{sdp}) implies that $F^{0,2}=0$, i.e.
the $A_{\zb_1}, A_{\zb_2}$ define a holomorphic structure on the
bundle ${\CE}_z = {\rm ker}{\CD}^{\dagger}_z$ over $X^{\circ}$. As
we have a unitary connection, $F^{2,0}=F^{0,2}=0$.

{}From (\ref{exct}) the holomorphic bundle ${\CE}$ extends to a
holomorphic bundle ${\CF}$ on  the whole of ${\R}^4$. We will now
construct a compactification $X$ of $X^{\circ}$ with a holomorphic
bundle ${\tilde\CE}$ over $X$ such that ${\tilde\CE}
\vert_{X^{\circ} } \approx {\CE}$, and whose connection ${\tilde
A}$ is a smooth continuation of the connection $A$ over
$X^{\circ}$. This compactification $X$ projects down to ${\C}^2$
via a map $p: X \to {\C}^2$. The pull-back $p^*{\CF}$ is a
holomorphic bundle over $X$ which differs from ${\tilde\CE}$. This
difference is localised at the exceptional variety, which is the
preimage $p^{-1}(Z)$ of the set of freckles.

\section{The Abelian case in detail}

Let us rotate $\vec \z$ so that ${\z}_{\C} = 0, \zeta_\R = \zeta>0$
and consider the case $w = 1$ or $U(1)$ or abelian instantons.  As we have
already remarked, Nakajima
\cite{nakajima} shows that $J = 0$. Hence,
$I^{\dagger}I = 2v\zeta$ and $[B_{1},B_{2}]=0$. When the matrices
$B_{1,2}$ and $I$ satisfy the stability criterion given earlier, the
moduli space we are describing is the Hilbert scheme $\Hilb_v(\C)$.
At the outset note that
abelian instantons do not exist for $\R\sp4$.

We can now solve the equations (\ref{ferm}) rather explicitly:
\begin{equation}\label{sli}
\begin{pmatrix}\Psi_{1} \\ \Psi_{2} \end{pmatrix}
= - \begin{pmatrix} B_{2}^{\dagger} -{\zb}_2 \\ B_{1}^{\dagger} -{\zb}_1
\end{pmatrix}
{\bx} I {\chi},
\end{equation}
where
\begin{equation}\label{bbx}
{\bx}^{-1} =  (B_{1}-z_1)(B_{1}^{\dagger}-{\zb}_1)
+ (B_{2}-z_2)(B_{2}^{\dagger}-{\zb}_2)
\end{equation}
and
\begin{equation}\label{cch}
\chi = \frac{1}{\sqrt{1 + I^{\dagger}{\bx} I}}.
\end{equation}
Let ${\bf P}(z) = {\rm Det}{\bf G}^{-1}$. It is a polynomial in
$z, {\zb}$ of degree $v$. Clearly (\ref{cch}) implies that: $$
{\chi}^{2} = \frac{{\bf P}(z)}{ {\bf Q}(z)} $$ where ${\bf Q}(z) =
{\bf P}(z) + I^{\dagger} {\widetilde{\bf G}^{-1}} I$ is another
degree $v$ polynomial in $z, {\zb}$, ${\widetilde{\bf G}^{-1}}$
being the matrix of minors of ${\bf G}^{-1}$.

The gauge field (\ref{gf}) is calculated to be
\begin{equation}\label{gage}
A = (\p - \pb) {\rm log} {\chi},
\end{equation}
and its curvature is
\begin{equation}\label{vti}
F = \p \pb {\rm log}{\chi}^2.
\end{equation}

The formula (\ref{gage}) provides a well-defined one-form on
the complement $X^{\circ}$
in ${\R}^4$ to the set $Z$ of zeroes of ${\bf P}(z)$.
This is just where $B_1-z_1$ and $B_2-z_2$ fail to be
invertible (and so $\sigma_z$ fails to be injective), that is a ``freckle".
Here we will only study the case of one point in detail and record that
the higher charge case can be dealt with similarly, with integrable systems
calculations helping greatly \cite{BN}.
The case of just one freckle already yields a surprise: abelian instantons
exist. Let us examine what is going on.

\subsection{Charge one instantons.}

To see what happens at such a point consider the case $v=1$.
Then (after shifting ${\zb}_1$ by $B_{1}^{\dagger}$,  {\it etc.})
\begin{equation}\label{fermi}
\Psi_{z} = \frac{1}{r\sqrt{r^2 + 2\zeta}}
\begin{pmatrix}{\zb}_{1} \sqrt{2\z}\\
{\zb}_{2} \sqrt{2\z} \\ r^2 \end{pmatrix}, \quad
\chi = \frac{r}{\sqrt{r^2 + 2\z}},
\end{equation}
where $r^2 = \vert z_1 \vert^2 + \vert z_2 \vert^2$.
Thus in this case
$$
{\bf P}(z) = z_1 {\zb}_1 + z_2 {\zb}_2, {\quad} {\bf Q}(z) =
z_1 {\zb}_1 + z_2 {\zb}_2 + 2{\z}
$$
The gauge field is given by (setting $2\zeta = 1$):
\begin{equation}\label{gge}
A = \frac{1}{2r^2 ( 1 + r^2)} \left(
z_1 {\dd} {\zb}_1  - {\zb}_1 {\dd} z_1 +
z_2 {\dd} {\zb}_2  - {\zb}_2 {\dd} z_2 \right),
\end{equation}
and
\begin{equation}\label{vta}
F =\frac{ {\dd}z_1\wedge {\dd}{\zb}_1+
{\dd}z_2\wedge {\dd}{\zb}_2 }{ r^2 ( 1+ r^2)}
-
\frac{1+2r^2}{r^4 ( 1+ r^2)^2}   \sum_{i,j} z_i {\zb}_j {\dd}z_j \wedge
{\dd}{\zb}_i .
\end{equation}

\subsection{The first blowup}

To examine (\ref{gge}) further let us rewrite $A$ as follows:
$$
A = A_{0} - A_{\infty},
$$
$$A_{0} =
\frac{1}{2r^2} \left(
z_1 {\dd} {\zb}_1  - {\zb}_1 {\dd} z_1 +
z_2 {\dd} {\zb}_2  - {\zb}_2 {\dd} z_2 \right),$$
$$A_{\infty} =
\frac{1}{2(1+r^2)} \left(
z_1 {\dd} {\zb}_1  - {\zb}_1 {\dd} z_1 +
z_2 {\dd} {\zb}_2  - {\zb}_2 {\dd} z_2 \right).$$
The form $A_{\infty}$ is regular everywhere in ${\R}^4$.
The form $A_{0}$ has a singularity at $r=0$. Nevertheless, as
we now show, this  becomes a well-defined gauge field on ${\R}^4$
blown up at one point $z = 0$.

Let us describe the blowup in some detail. We start with ${\C}^2$
with coordinates  $(z_1, z_2)$. The space blown up at the point
$0 = (0,0)$ is simply the space $X$
of pairs $(z,{\ell})$, where
$z \in {\C}^2$, and ${\ell}$ is a complex line which passes through
$z$ and the point $0$. $X$ projects to ${\C}^2$ via the map
$p (z, {\ell}) = z$. The fiber over each point $z \neq 0$
consists of a single point while the fiber over the  point $0$
is the space $\CP1$ of complex lines passing through
the point $0$.

In our applications we shall need a coordinatization of the blowup.
The total space of the blowup is a union
$X = {\CU} \cup {\CU}_{0} \cup {\CU}_{\infty}$ of three
coordinate patches. The local coordinates in the patch ${\CU}_0$
are $(t, {\l})$ such that
\begin{equation}\label{op}z_1 = t, \, z_2 = {\l}t.\end{equation}
In this patch ${\l}$ parameterises
the complex lines passing through the point $0$, which are not
parallel to the $z_1 = 0$ line.
In the patch ${\CU}_{\infty}$ the coordinates  are
$(s ,{\mu})$, such that
\begin{equation}\label{anp}
z_1 = {\mu}s, z_2 = s.
\end{equation}
There is also a third patch ${\CU}$, where $(z_1, z_2) \neq 0$.
This projects down to ${\C}^2$ such that over each point
$(z_1,z_2) \neq 0$ the fiber consists of just one point. The fiber
over the point $(z_1, z_2) =0$ is the projective line
$\CP1 = \{ {\l}  \} \cup \infty$. We now show that on this
blown up space our gauge field is well defined.

On ${\CU} \cap {\CU}_{0}$ we may write
\begin{equation}\label{ggei}
A_{0} = {\frac{t {\dd} {\tb} - {\tb} {\dd} t}{2\vert t \vert^2}} +
{\frac{{\l} {\dd} {\bar{\l}} - {\bar {\l}}{\dd} {\l}}
            {2(1+\vert {\l} \vert^2)}}.\end{equation}
Define $A_{{\CU}_{0, \infty}}$ as
\begin{equation}
\label{ggeo}
\begin{split}
A_{{\CU}_{0}} & =
\frac{{\l} {\dd} {\bar{\l}} - {\bar {\l}}{\dd} {\l}}
     {2(1+\vert {\l} \vert^2)},\\
A_{{\CU}_{\infty}} &=
\frac{{\mu} {\dd} {\bar{\mu}} - {\bar {\mu}}{\dd} {\mu}}
     {2(1+\vert {\mu} \vert^2)}.
\end{split}
\end{equation}
Now $A_{0}$ is a well-defined one-form on $\CU$.
On the intersections ${\CU} \cap {\CU}_{0}$
the one-forms $A_{0}$ and $A_{{\CU}_{0}}$ are related via
a gauge transformation $$i \,{\dd}\, {\rm arg} t .$$
On the intersection ${\CU}_{0} \cap {\CU}_{\infty}$
the one-forms $A_{{\CU}_0}$ and $A_{{\CU}_{\infty}}$
are related via $$i\, {\dd} \,{\arg} {\l} = - i\, {\dd} \, {\rm arg} {\mu}$$
gauge transformations. Finally on ${\CU} \cap {\CU}_{\infty}$
the one-forms $A_{0}$ and $A_{{\CU}_{\infty}}$
are related via the gauge transformation
$$i \,{\dd} \,{\rm arg} s .$$
We have shown therefore that $A_0$ is a well-defined gauge field on
$X$. Observe also that at infinity $A \to 0$ as $o(r^{-3})$, which
yields a finite action. In fact
the gauge field (\ref{gge}) has a non-trivial Chern class ${\rm ch}_2$:
\begin{equation}\label{acti}
F \wedge F =  - \frac{2}{r^2 (1 + r^2)^3} {\dd}z_1 \wedge
{\dd}{\zb}_1 \wedge {\dd}z_2 \wedge {\dd}{\zb}_2
\end{equation}
so that
$$
\frac{1}{4\pi^2} \int F \wedge F = 1.
$$
Finally, the restriction of $A$ on the exceptional divisor $E$,
defined by the equation $t = 0$ in ${\CU}_0$ and $s = 0$ in ${\CU}_{\infty}$,
has non-trivial first Chern class:
$$
\frac{1}{2\pi i} \int_{E} F = - 1.
$$

The reason that an abelian instanton exists is that space-time is blown up,
and now there are noncontractible $2$-spheres. The space under discussion
is not $\C\sp2$. We should also remark that since the curvature of our
gauge field has type the $(1,1)$ and is non-degenerate on the blowup
it can be used as a K\"ahler form. Then, tautologically, the gauge field has
the  same self-duality property as the K\"ahler form (it is either self-dual
or anti-self-dual depending on the choice of orientation). The complex
coordinates for this are in general complicated expressions of the
coordinates $z_{1,2}$ we have employed.

\subsection{Comparison with the Born-Infeld instanton}
Notice the similarity of the solution (\ref{gge}) to the formulae
(4.56), (4.61) of the paper \cite{witsei}. It has the same
asymptotics both in the $r^2 \to 0$ and $r^2 \to \infty$   limits.
Of course the formulae in \cite{witsei} were meant to hold only
for slowly varying fields and that is why we don't get precise
agreement. Nevertheless, we conjecture that all our gauge fields
are the transforms of the non-commutative instantons from
\cite{neksch} under the field redefinition described in
\cite{witsei}. {}From our analysis it follows that one has
to modify the topology of space in order to make non-singular the
corresponding gauge fields of the ordinary gauge theory.

\subsection{Comparison with the non-commutative instanton}

It is instructive to compare the solutions above with those
defined on the noncommutative space. Traditionally one describes
the moduli space with non-vanishing parameters ${\zeta}$ as
corresponding to the instantons on noncommutative ${\R}^4$ where
all four coordinates are noncommuting. In order to make these
solutions as close to the commutative ones as possible we shall
consider the four dimensional space which is a product of the
ordinary commutative plane, with the coordinates $z, {\zb}$ and
the noncommutative plane, i.e. the algebra $[a, a^{\dagger} ] =
1$. This would correspond to the parameters ${\zeta}_{\C} = 0$ and
${\zeta}_{\R} = 1$.  For simplicity, set $B_2 = 0$, that is,
consider the instantons, elongated at $a^{\dagger} a = 0$, along the
$z$ direction. Then the application of the ADHM construction
yields the following formulae for the elongated instantons on this
space:
\begin{eqnarray}
\label{eq:noncm} & A = {\xi}^{-1} {\pb} {\xi}  - {\p}
{\xi}^{\dagger} {\xi}^{\dagger \ -1} \nonumber \\ & {\xi} = {\xi}
(z, a^{\dagger}a ), \qquad {\xi} (z , n) = {\xi}_{n}(z) \nonumber
\\ & {\xi}_n (z) = \sqrt{\frac{\Det \left( (B_1 - z)(B_1^{\dagger} -
{\zb}) + n \right)} {\Det \left( (B_1 - z)(B_1^{\dagger} -
{\zb}) + n +1 \right)}}, \quad n > 0 \nonumber \\ & {\xi}_0 (z) =
{\frac{\Det (B_1 - z) }{\sqrt{{\Det} \left( (B_1 -
z)(B_1^{\dagger} - {\zb}) + 1 \right)}}} \qquad
\end{eqnarray}
Here, ${\pb} = d{\zb} {\frac{\p}{{\p}{\zb}}} + \ da \ [ \cdot,
a^{\dagger}]$, and $a^{\dagger}a = n$. The solution
(\ref{eq:noncm}) is non-singular, without any topology change.
However, there is a noncommutative indication of the blowup. It is in
the phase of the function ${\xi}$. For $n > 0$ it vanishes, while
for $n=0$ it is winding around the zeroes of ${\Det} ( B_1 - z)$.
In the commutative description this would have been described as
the local gauge transformation, patching the regions with
$a^{\dagger} a \sim 0$ and those with $a^{\dagger} a \gg 0$. The
winding of this gauge transformation $$ {\exp} \ i \ {\rm arg}
{\Det} (B_1 - z)$$ is related with the number of the points, blown
up in the commutative description.

\section{The Burns metric}
We shall now show that there is a particularly nice metric on the
blow-up of $\C\sp2$ for which our charge one instanton is
anti-self-dual. This is the Burns metric \cite{Burns} which is
scalar flat with \asd Weyl curvature  $W\sp+=0$. We remark that
$F\sp+=0$ is the correct equation to go with $W\sp+=0$ if there is
to be twistor correspondence.

Consider the K\"ahler form on ${\mathbb C}\sp2 -\{0\}$,
$$ \Omega= -{\frac{i}{2}}\p\pb\left( |{\bf z}|\sp2 + m\log |{\bf z}|\sp2
\right).$$
We have a volume element
$$\frac{1}{2}\Omega\wedge  \Omega=-\frac{1}{4}\left(1+\frac{m}{r\sp2}\right)
dz_1\wedge d\bar z_1\wedge dz_2\wedge d\bar z_2 .$$
Let $r^2=|{\bf z}|\sp2=z_1\bar z_1 +z_2\bar z_2$.
Then with $g(Ix,y)=\Omega(x,y)$ and $Idz_1=idz_1$ etc we get

$$g_{1\bar1}=g_{\bar1 1}=
  {\frac{1}{2}} \left(1+{\frac{m}{r\sp2}}-{\frac{m |z_1|\sp2}{r\sp4}} \right)$$
$$g_{2\bar2}=g_{\bar2 2}=
  {\frac{1}{2}} \left(1+{\frac{m}{r\sp2}}-{\frac{m |z_2|\sp2}{r\sp4}} \right)$$
$$g_{1\bar2}=g_{\bar2 1}= -{\frac{m}{2 r\sp4}}z_2\bar z_1$$
$$g_{2\bar1}=g_{\bar1 2}=-{\frac{m}{2 r\sp4}}z_1\bar z_2$$ For
$m=0$ this gives us the usual flat conventions. With
$\phantom{.}\sp\star \Omega= \Omega$ we have $\{\Omega, dz_1\wedge
dz_2, d\bar z_1\wedge d\bar z_2\}\in \Lambda_+\sp2 T\sp\star M$.

Now
$$\Lambda\sp2 T\sp\star M=
  \Lambda_+\sp2 T\sp\star M\oplus \Lambda_-\sp2 T\sp\star M.
$$ and $\Lambda_-\sp2 T\sp\star M$ consists of the $(1,1)$ forms
orthogonal to $\Omega$. With $$\Omega\wedge\star\alpha
=-\frac{1}{2}(\Omega,\alpha )\ \Omega\wedge  \Omega$$ we see that
if $F\wedge  \Omega=0$ then $F\in \Lambda_-\sp2 T\sp\star M$. We
have explicitly calculated the abelian instanton. In terms of
$\chi$ determined from the ADHM data we have $$ {A = (\p - \pb)
{\rm log} {\chi}}, $$ and $$ {F = \p \pb {\rm log}{\chi}^2.} $$ With
$$ \chi = {\frac{r}{\sqrt{r^2 + m}}}, $$ (here $m=2\zeta$ in our
notation) we find $F\wedge  \Omega=0$. Thus $F=\phantom{.}\sp\star
F$ with this metric. Observe that ${\rm Ricc}\wedge\Omega=0$ and
so our metric is self dual as stated. For both these calculations
it is convenient to note $$\p\pb\,f(r\sp2) \wedge \p\pb\,h(r\sp2)=
2\,r\sp2 h' f' \left(\frac{f''}{f'}+\frac{h''}{h'}+\frac{2}{r\sp2}
\right)\, \star 1 $$
where $r\sp2=z_1\bar z_1+z_2\bar z_2$ and $f'={\d
f(x)}/{\d x}$ and so forth.

\section{Discussion}
Thus far we have identified the phase space of the ( complexified
$a_n$) Calogero-Moser systems with the moduli spaces of deformed
instantons and that of instantons on a non-commutative space-time.
We have shown that there is a very nice metric, the Burns metric,
on the one-instanton space-time for which these deformed instanton
equations are in fact anti-self-dual, i.e. solve the ordinary
instanton equation. Thus far our discussion has focused on the
real structure of these spaces, and as real spaces they are
diffeomorphic. There is more to the story however. These spaces
have complex strucures, and as we described earlier, these are
different. A choice of complex structure (or a $B$-field) effects
this description. We will conclude by briefly recording some of
these differences.

We have already described the Hilbert scheme of points in terms of
codimension $v$ ideals in $A_0=\C[x,y]$. The Calogero-Moser phase space has
a description in terms of ideals in the $1$-st Weyl algebra $A_1$. The
crucial difference here is that $A_1$ has no finite dimensional
representations (for $\Tr([L,X]-{\bf 1}_V)=0$ yields $v=0$). However by
letting ${\cal C}_v$ be the space for which (up to conjugation)
$\{ \rank([L,X]-{\bf 1}_V)\}$ has at most $1$, Berest and Wilson \cite{BW} show
${\cal C}={\bigsqcup}_{v\ge 0}\,{\cal C}_v$ is equivalent to the isomorphism
classes of right ideals in $A_1$. Further Ginzburg has shown there is an
infinite algebraic group $G_1$ acting on ${\cal C}$ such that it acts transitively
on ${\cal C}_v$. Thus ${\cal C}_v$ is the coadjoint orbit of $G_1$. This
parallels the results that hold for the Hilbert scheme. These similarities
and difference are summarized in the following:
$$
\begin{array}{cc}
{\rm Calo}_v&\Hilb_v\\ \\
\begin{array}{rl}
\mu_\C=&[L,X]+u\sp{T} u=\z_\C {\bf 1}_V \\
\mu_\R=&0\\
\\
A_1=&\frac{\C<x,y>}{[x,y]-1}\\
\\
{\cal C}_v=&\frac{\{ \rank([L,X]-{\bf 1}_V)\le 1\}}{\sim}\\
\\
{\cal C}=&{\bigsqcup}_{v\ge 0}\,{\cal C}_v\\
\\
&G_1\ {\rm generated\ by}\\
\Phi_p(x)&=x-p'(y)\qquad \Phi_p(y)=y\\
\Phi_q(x)&=x\qquad \Phi_q(y)=y+q'(x)\\
&p,q\ {\rm polynomials}
\end{array}
&
\begin{array}{rl}
\mu_\C=&[B_1,B_2]+IJ=0, \quad J=0\\
\mu_\R=&\z_\R {\bf 1}_V\\
\\
A_0=&\C[x,y]\\
\\
{\cal D}_v=&\frac{\{ [B_1,B_2]=0, {\rm cyclic vector}\}}{\sim}\\
\\
{\cal D}=&\bigsqcup_{v\ge 0}\,{\cal D}_v\\
\\
&G_0\ {\rm generated\ by}\\
&\Phi_{lin}(x)=ax+by\qquad \Phi_{lin}(y)=cx+dy,\\
&\Phi_{jonq}(x)=x+p(y)\qquad\ \ \Phi_{jonq}(y)=y\\
&p\ {\rm polynomial}, \ \ ad-bc=1
\end{array}\\
\end{array}
$$

We conclude by stressing that at the moment the physical, D-brane
interpretation of the Burns metric is far from being clear. The
``commutative'' description of the noncommutative
instantons, following from the analysis of the Dirac-Born-Infeld
action for the gauge theory on the D-brane in the presence of
background magnetic field is usually performed in the static
gauge. The singularity of the naive expressions for the gauge
field may signal the invalidity of this gauge, suggesting that the
topology of the worldvolume of the brane (not of the ambient flat
ten dimensional space-time) is non-trivial.

\section*{Acknowledgements}
We have benefited from discussions with M.F. Atiyah, D.J.M.
Calderbank, V. Fock, A.~Gorsky, A.~Marshakov, A.~Mironov,
A.~Morozov, M.A.~Olshanetsky and V. Rubtsov. H.W.B. is grateful to the
organizers of these workshops for the stimulating atmosphere fostered.

\end{document}